\documentclass[acmsmall,screen,nonacm]{acmart}

\setcopyright{none}
\renewcommand\footnotetextcopyrightpermission[1]{}

\usepackage{amsmath}
\usepackage{array}
\usepackage{booktabs}
\usepackage{multirow}
\usepackage{siunitx}
\usepackage{tabularx}
\usepackage{xspace}

\newcolumntype{Y}{>{\raggedright\arraybackslash}X}

\newcommand{\qtwofive}{Qwen2.5-Coder-7B\xspace}
\newcommand{\qthreefive}{Qwen3.5-9B\xspace}
\newcommand{\pp}{\,pp\xspace}
\newcommand{\methodone}{One-shot\xspace}
\newcommand{\methodlegacy}{Evolve\xspace}
\newcommand{\methodmatched}{Matched\xspace}

\AtBeginDocument{%
  }

\begin{document}

\title{Auditing and Decomposing Feedback-Driven Evolution in LLM Test Generation under the Oracle Problem}

\author{Yunhao Liang}
\affiliation{%
  \institution{Chengdu Institute of Computer Applications, Chinese Academy of Sciences}
  \country{China}}
\affiliation{%
  \institution{University of Chinese Academy of Sciences}
  \country{China}}

\author{Chengguang Gan}
\affiliation{%
  \institution{Independent Researcher}
  \country{Japan}}

\author{Ruixuan Ying}
\affiliation{%
  \institution{Institute of Multidisciplinary Research for Advanced Materials (IMRAM), Tohoku University}
  \country{Japan}}

\author{Hanjun Wei}
\affiliation{%
  \institution{University of Chinese Academy of Sciences}
  \country{China}}

\author{Zhe Cui}
\affiliation{%
  \institution{Chengdu Institute of Computer Applications, Chinese Academy of Sciences}
  \country{China}}
\affiliation{%
  \institution{University of Chinese Academy of Sciences}
  \country{China}}

\author{Shiwen Ni}
\affiliation{%
  \institution{Artificial Intelligence Research Institute,\linebreak[1] Shenzhen University of Advanced Technology}
  \country{China}}

\renewcommand{\shortauthors}{Liang et al.}
\authorsaddresses{}

\begin{abstract}
Execution feedback can make LLM test generation appear self-verifying even when
generated inputs or outputs are invalid.  A common shortcut executes each input
on one accepted program and treats its output as ground truth, converting
out-of-domain behavior into apparent fault detection and evolutionary progress.

We audit this failure mode on 142 development, 114 locked external, and 138
held-out but qualification-amended tasks under two models, three seeds, and fault-cross-fitted
real submissions.  On external inputs where three accepted implementations
agree, generated outputs match the panel only 27.79\% and 50.12\% of the time.
A single reference inflates the apparent evolutionary gain by 9.46--14.85
percentage points; after audit, equal-budget independent resampling is
6.01--18.83 points better.

We then compare a genuine three-round loop with a density-matched feedback
placebo.  External Real--Placebo contrasts are $+0.13$ and $-0.50$ points.  An
outcome-blind held-out follow-up, transparently amended from 20 to 18 faults
after its original qualification gate stopped before model calls, yields $+1.99$ (95\% CI
[$+0.08$,$+3.88$]) and $+0.28$ points ([$-1.41$,$+2.03$]); neither meets the
pre-specified equivalence or superiority rule within the amended cohort.  Post-hoc matching of observed
reference-valid yield changes them to $-0.30$ and $+0.78$ points, both with
zero-crossing intervals.  The positive Qwen2.5 contrast is therefore not robust
to usable-candidate opportunity.  Because the external placebo is post hoc and
held-out qualification was amended, RQ6 supplies diagnostic boundary evidence,
not a pristine confirmation.  A blinded human semantic audit conducted by two
software engineering doctoral students jointly classifies 94.41\% of
panel-disconfirmed inputs as invalid but 3.60\% as valid, making
panel disagreement informative rather than semantic proof.  We provide an
audit-and-placebo protocol that separates verifier artifacts, interaction
scaffolding, and grounded feedback credit.
\end{abstract}

\ccsdesc[500]{Software and its engineering~Software testing and debugging}

\keywords{LLM-based test generation, self-evolution, test oracle problem,
differential testing, empirical software engineering}

\setlength{\emergencystretch}{2em}
\maketitle
\setlength{\emergencystretch}{0pt}
\thispagestyle{plain}
\pagestyle{plain}

\section{Introduction}
\label{sec:introduction}

Executable tests are attractive feedback for code models: they expose concrete
failures and appear to turn generation into a verifiable objective.  Recent
systems use generated tests to rank programs~\cite{chen2023codet}, refine suites
from execution feedback~\cite{cai2026codecontestso}, or co-train code and test
generators~\cite{wang2025cure,lee2026utrl}.  Yet execution verifies a generated
test only if its input is permitted by the natural-language specification and
its expected output is correct.  This is the test-oracle problem in a new
feedback loop~\cite{barr2015oracle}.

Competitive-programming data offer a tempting shortcut: execute a generated
input on one accepted program and treat its output as ground truth.  That
program may accept an out-of-domain input, fail on a valid corner case, or
choose behavior the specification leaves undefined.  Optimization can then
reward tests that disagree with many wrong submissions for the wrong reason.
The risk is amplified in self-evolution, where noisy feedback is repeatedly
selected, distilled, or reinforced.  More search can discover more
verifier-specific inputs and masquerade as learning unless the oracle and the
search budget are independently controlled.

We audit this failure mode in natural-language-to-test generation.  A model
receives a problem statement and up to three public examples and emits complete
input--output tests.  We compare one call, one call plus six deterministic
mutations, three independent calls with the same nine candidate slots, and
mutations of public inputs.  A deterministically chosen accepted program builds
the initial oracle.  Real faulty submissions are split so that one fold selects
an archive and the other evaluates it.  Only after selection is frozen do the
remaining accepted programs audit each selected test; disconfirmed entries are
removed without replacement.

We also evaluate a genuine three-round loop.  Later rounds receive prior inputs
and opaque identifiers for search-fold faults detected or remaining, never
held-out behavior.  Besides independent resampling, we introduce a
density-matched placebo that preserves the iterative prompt, history size,
detected-set cardinality, and kill-count density while randomizing the
input--fault correspondence.  Real--Placebo therefore tests whether
behavior-aligned feedback adds value beyond iterative scaffolding and coarse
progress.

The study covers 142 development tasks, a procedure-locked 114-task
TestCase-Eval cohort~\cite{yang2025testcaseeval}, and a 138-task held-out but
qualification-amended follow-up, under two models and three seeds.  It further decomposes 2,836 real
faults and conducts a blinded, stratified human semantic audit of 300
panel-disconfirmed inputs with two software engineering doctoral reviewers.
Tasks, not individual tests or executions, are the unit of inference.

Single-reference evaluation inflates the estimated evolution gain by
9.46--14.85\pp.  After audit, equal-budget resampling is 6.01--18.83\pp better
than mutation-based evolution in every model--cohort condition.  In the
three-round study, the large model-dependent Plain--iterative shifts are almost
reproduced by the placebo.  On the held-out qualification-amended cohort, Real--Placebo is
$+1.99$\pp (95\% CI [$+0.08$,$+3.88$]) and $+0.28$\pp
([$-1.41$,$+2.03$]); neither meets the frozen equivalence or practical-
superiority rule.  Yield-matched replay changes the contrasts to $-0.30$ and
$+0.78$\pp, both with zero-crossing intervals.  Fine-grained feedback credit is
therefore unresolved, not established absent.  The semantic audit jointly
judges 94.41\% of panel-disconfirmed inputs invalid but 3.60\% valid, so panel
disagreement remains evidence rather than semantic proof.

This paper contributes:

\begin{itemize}
  \item \emph{Oracle-induced gain inflation} and a freeze-then-audit protocol
  using independently submitted accepted implementations.

  \item Candidate-budget-matched, fault-cross-fitted experiments on 394 tasks,
  including a three-round execution-feedback loop under strict external
  execution isolation.

  \item A density-matched feedback placebo, an outcome-blind but
  qualification-amended held-out follow-up, and an explicit evidence-status
  account separating interaction scaffolding from behavior-aligned attribution.

  \item Mechanism evidence from 2,836 real faults and a hash-locked human
  semantic audit of 300 stratified inputs, together with
  task-level evidence for independent reconstruction.
\end{itemize}

The contribution is diagnostic rather than a new generator: feedback validity,
equal-budget search, and a suitable placebo must be established before a gain
can be attributed to evolution or execution feedback.

\section{Background and Problem Formulation}
\label{sec:background}

\subsection{Tests as feedback}

For a programming task $q$, let $S_q^+$ denote independently submitted programs
that are accepted by the benchmark and $S_q^-$ denote historical programs with
wrong verdicts.  A generated test $t=(x,y)$ contains an input $x$ and an
expected output $y$.  A suite is useful only if it accepts specification-conforming
programs and rejects faulty ones.  Because membership in $S_q^+$ is a platform
verdict rather than a proof, the following fidelity is operational with respect
to the accepted-program panel.  We use the conventional fidelity and fault-detection
quantities
\begin{align}
 \operatorname{Fid}_q(T)
   &= \frac{1}{|S_q^+|}\sum_{s\in S_q^+}
      \mathbf{1}[s \text{ passes } T], \\
 \operatorname{Kill}_q(T)
   &= \frac{1}{|S_q^-|}\sum_{s\in S_q^-}
      \mathbf{1}[s \text{ fails } T].
\end{align}
The accepted-panel rejection rate is $1-\operatorname{Fid}_q(T)$.  Kill rate is
meaningless without fidelity because an arbitrary expected output can reject
every program.  Unlike mutation testing~\cite{jia2011mutation}, our faults are
real historical submissions.

\subsection{Three notions of oracle validity}

We separate three checks that are often conflated.

\paragraph{Direct model oracle.}
For model output $(x,y_m)$, $x$ is \emph{panel-admissible} when every available
$s\in S_q^+$ completes and their normalized outputs agree on $y_A(x)$.  A direct
oracle match is $y_m=y_A(x)$.  This post-generation diagnostic never affects
selection and measures panel agreement, not semantic truth.

\paragraph{Single-reference oracle.}
A source hash selects $r_q\in S_q^+$ before generation.  If $r_q(x)$ completes,
the primary verifier sets $y_r=r_q(x)$ and discards the generated output.  This
removes direct output hallucination but does not validate $x$.

\paragraph{Multi-implementation audit.}
After freezing the primary archive, every remaining $S_q^+$ program runs each
selected input.  A test is panel-confirmed only if all complete and agree with
$y_r$; otherwise it is \emph{audit-disconfirmed} and removed without
replacement.  The label denotes differential disagreement, not semantic
falsity: accepted programs may contain or share latent defects.

\subsection{Oracle-induced gain inflation}

Let $K_1(M)$ be method $M$'s held-out kill under one reference and $K_A(M)$ the
score of the same frozen archive after panel audit.  For evolution $E$ and its
one-shot parent $O$,
\begin{equation}
  I(E,O) = [K_1(E)-K_1(O)]-[K_A(E)-K_A(O)].
  \label{eq:inflation}
\end{equation}
Positive $I$ is gain that does not survive audit.  Comparing $E$ with
equal-slot resampling $R$ under $K_A$ separates the operator from additional
search.

\subsection{Attributing value to execution feedback}

A multi-round interface also adds an iterative prompt, history, and coarse
progress.  \emph{Real feedback} maps each input to the search faults it detects;
\emph{placebo feedback} preserves interaction structure and density but destroys
that mapping.  For audited real and placebo scores $K_A(F)$ and $K_A(P)$,
\begin{equation}
  A(F,P) = K_A(F)-K_A(P)
  \label{eq:aligned-feedback}
\end{equation}
measures aligned attribution. The placebo retains aggregate progress. Its
contrast with resampling therefore mixes scaffolding and coarse feedback. A
zero-crossing interval detects no benefit but does not establish equivalence.

\subsection{Scope of ``evolution''}

Self-evolving systems can change weights, memory, tools, or artifacts
~\cite{gao2025selfevolving}.  We study artifact-level evolution: deterministic
candidate mutation and a three-round loop in which search-fold outcomes affect
later prompts.  This stage could supply data or reward to later training, but
we do not claim parameter-level self-improvement.

\section{Study Design}
\label{sec:design}

We ask six questions: \textbf{RQ1}, how often model-generated outputs agree
with the accepted-program panel; \textbf{RQ2}, how much one reference inflates
evolution gains; \textbf{RQ3}, whether audited evolution beats one shot or
equal-slot resampling; \textbf{RQ4}, which fault families concentrate
disconfirmed detections and whether rankings are robust; \textbf{RQ5}, whether
blinded judgments support the invalid-input mechanism; and \textbf{RQ6}, what
incremental value behavior-aligned feedback shows beyond a density-matched
iterative placebo under exposed and qualification-amended cohorts.

\subsection{Tasks and model conditions}

Table~\ref{tab:cohorts} summarizes three non-pooled cohorts. Development uses
142 exact-output CodeContests-family tasks previously exposed during method
development. The external cohort was hash-locked before fresh generation from
TestCase-Eval~\cite{yang2025testcaseeval}: 114 non-interactive tasks with a
complete statement, public examples, three accepted Python programs passing
those examples, and 20 nonoverlapping faulty programs. The held-out cohort
comes from a 200-task CodeContests test bundle closed during earlier studies.
Its frozen gate required 20 faults and at least 40 tasks per source, but stopped
before model calls because only 36 OOD tasks qualified. A separately frozen,
outcome-blind amendment changed only the fault panel and folds from 20/10--10 to
18/9--9, retaining 71 IID and 67 OOD tasks. We report this cohort as
qualification-amended, not pristine; the complete flow is in the supplemental
material. Python and exact-output tasks provide common, resource-limited
execution semantics and avoid special-checker ambiguity.

\begin{table}[t]
  \caption{Cohorts are analyzed separately. Counts are per task.}
  \label{tab:cohorts}
  \centering
  \small
  \begin{tabular}{lrrrrl}
    \toprule
    Cohort & Tasks & Accepted & Faulty & Seeds & Role \\
    \midrule
    Development IID & 63 & 9--10 & $\leq20$ & 3 & Development \\
    Development OOD & 79 & 3--10 & $\leq20$ & 3 & Development \\
    External & 114 & 3 & 20 & 3 & Procedure-locked evaluation \\
    Held-out E16-B & 138 & 3 & 18 & 3 & Qualification-amended \\
    \bottomrule
  \end{tabular}
\end{table}

The RQ6 evidence has deliberately different status across phases
(Table~\ref{tab:rq6-status}). ``Locked'' therefore refers to the stated
procedure or decision rule, not automatically to a pristine confirmatory
sample.

\begin{table}[t]
  \caption{RQ6 evidence status. No single phase is treated as a pristine
  confirmation of the fine-grained feedback effect.}
  \label{tab:rq6-status}
  \centering
  \small
  \begin{tabularx}{\textwidth}{lYYY}
    \toprule
    Study & Timing and cohort & Departure from a pristine test & Evidential role \\
    \midrule
    E14 & Procedure locked before generation on exposed External tasks & Cohort already exposed & Prospective extension \\
    E15 & Constructed after E14 on the same tasks & Post-hoc placebo & Diagnostic control \\
    E16-A & Gate locked before opening 200 held-out tasks & Qualification failed; no model calls & Stopped before generation \\
    E16-B & Frozen before model outcomes & Fault threshold amended after supply was observed & Outcome-blind boundary evidence \\
    Replay & Constructed after E16-B outcomes & Conditions on observed candidate yield & Sensitivity analysis \\
    \bottomrule
  \end{tabularx}
\end{table}

The first model condition is a \qtwofive Instruct checkpoint
~\cite{hui2024qwen25coder} merged with a fixed warm start (49 records, 33 tasks,
LoRA rank 8, three epochs). The second is the released \qthreefive checkpoint
with thinking disabled~\cite{qwen2026qwen35}. They test replication, not a
causal scale effect. For seeds 42--44, each receives the same statement and up
to three public examples and returns at most three strict JSONL tests.
Generation uses temperature 0.7, top-$p$ 0.8, top-$k$ 20, repetition penalty
1.1, a 1,024-token completion limit, and a 4,096-token context.

\subsection{Candidates, cross-fitting, and auditing}

All primary branches share the first call (Table~\ref{tab:methods}).
\methodone retains its at-most-three parsed inputs. \methodlegacy adds six
deterministic mutations without observing fault outcomes. \methodmatched adds
two calls under the identical prompt. Public evolution applies nine mutations
to public inputs and uses no model call. Failed mutations are not resampled,
and every archive contains at most three tests.

\begin{table}[t]
  \caption{Per-task-repeat candidate budgets.}
  \label{tab:methods}
  \centering
  \small
  \begin{tabular}{lrrl}
    \toprule
    Variant & Calls & Slots & Additional candidates \\
    \midrule
    \methodone & 1 & 3 & None \\
    \methodlegacy & 1 & 9 & 6 deterministic mutations \\
    \methodmatched & 3 & 9 & 2 independent calls \\
    Public evolve & 0 & 9 & 9 public-input mutations \\
    \bottomrule
  \end{tabular}
\end{table}

Generated outputs are used only for RQ1. For selection, a minimum-source-hash
accepted program executes each unique input and supplies its expected output;
failures are rejected. A content hash splits faults into balanced folds. In
each direction, greedy selection covers one search fold, while the other fold
evaluates the at-most-three-test archive; the roles then swap.

The archive is frozen before the other accepted programs execute its inputs.
An entry is removed without replacement if any auditor fails or disagrees.
Thus single-reference and audited scores describe the same selected archive.
The primary outcome is task-macro held-out constrained kill. Secondary outcomes
are whether any held-out fault is detected and whether any test survives audit.
We call the latter \emph{audited task coverage}; it is the fraction of tasks
retaining a valid test, not source-code line or branch coverage. Seeds are
averaged within task before inference.

\subsection{Real and placebo feedback}

RQ6 compares three-call-per-direction interventions. Real feedback starts from
the shared first call; each of two later rounds sees bounded prior inputs,
opaque identifiers for search faults detected per input, and remaining search
fault identifiers. It never sees sources, outputs, stack traces, held-out
behavior, or audit outcomes. The placebo uses the same template, rounds,
history size, detected-set cardinality, and multiset of the two largest
per-input kill counts, but randomly reassigns history inputs and fault
identifiers. Independent resampling is the no-feedback baseline.

E14 locked the real procedure on the already exposed external cohort; E15's
placebo was designed after E14 and is post hoc. E16-B generated Plain, Real,
and Placebo afresh on 138 held-out tasks. Per task-repeat, Plain uses three
calls; Real and Placebo each share round zero across directions and use four
direction-specific follow-ups. Real and Placebo are therefore matched in both
calls and slots, whereas Plain is slot- but not unique-call-matched. Across
both models E16-B required 9,108 calls.

Nominal slots can fail parsing, duplicate an input, or fail reference execution.
After the frozen analysis, an off-policy replay retained, within every task,
seed, direction, and round, the earliest equal number of reference-valid Real
and Placebo candidates and repeated selection and audit. A stricter diagnostic
matched confirmed and disconfirmed yields separately. These post-treatment
analyses test sensitivity to usable opportunity; they are not counterfactual
reruns because later generations retain their original histories.

\subsection{Mechanism and semantic sensitivity}

For RQ4, each of 2,836 development faults is assigned one mutually exclusive,
execution-grounded trigger phenotype: public example, official exact-output
anchor, safe local mutation family, unsafe-public-mutation only, or hard
residual. The categories identify triggers, not source-level causes. Anchors
are re-executed across accepted programs and family intervals remain
exploratory.

RQ5 samples 300 of 1,110 deduplicated, selected, reference-executable but
panel-disconfirmed inputs using locked strata (150 development, 150 external;
construction and disagreement-mode weights). Two software engineering doctoral
students independently review each case using the statement, public examples,
and candidate input. They are blinded to task identity, generation method,
archived output, program behavior, fault detections, panel outcome, and the
other reviewer's judgment. Each chooses one of five labels: valid, out of
domain, malformed, ambiguous, or insufficient information. For a valid input,
the reviewer also independently derives an expected output. We report exact
inter-rater agreement
and Cohen's $\kappa$~\cite{cohen1960coefficient}, with design-weighted estimates
and 10,000 within-stratum bootstraps. These are structured human judgments, not
definitive semantic ground truth. A 67-case packet (all reviewer disagreements
and jointly valid cases plus consensus quality controls) is reserved for
consensus or third-reviewer specification adjudication.

\subsection{Inference and integrity}

Primary contrasts use paired task-level nonparametric bootstraps with 20,000
resamples and 95\% intervals. External gates were frozen: gain inflation must
have a positive lower bound, and audited \methodlegacy--\methodmatched must
have an upper bound no greater than $+2$\pp. E15 had no equivalence margin, so
zero-crossing intervals mean no detectable incremental effect, not equivalence.
E16-B pre-specified $\pm2$\pp as the smallest effect of interest: equivalence
requires the whole interval inside that range, superiority a lower endpoint
above $+2$\pp, and inferiority an upper endpoint below $-2$\pp, separately for
both models.

External programs run in bubblewrap namespaces with time, memory, process, and
output limits; unsafe fallback is forbidden. The older development runner
saved limits and counts but lacks equivalent per-row isolation attestations.
Model artifacts, cohorts, folds, generated rows, analyses, and paper summaries
are linked by SHA-256 locks. The failed E16-A gate and E16-B amendment are
separately signed.

\section{Results}
\label{sec:results}

\subsection{RQ1: Direct expected outputs remain unreliable}

Table~\ref{tab:oracle} evaluates only generated inputs on which all available
benchmark-accepted implementations execute and agree.  Even under that favorable
conditioning, the model-proposed expected output matches the panel for 44.16\% of
admissible development inputs in the \qtwofive condition and 58.90\% in the
\qthreefive condition.  The prospectively locked external rates are lower:
27.79\% (95\% CI [22.93, 32.89]) and 50.12\% [43.81, 56.43].  Requiring every
test in a call to be panel-confirmed is more demanding.  Fully panel-confirmed suites occur in only
2.53\% and 5.07\% of external calls.

\begin{table}[t]
  \caption{Direct-oracle reliability. ``Admissible'' counts tasks with at least
  one panel-admissible input. Rates and 95\% CIs are
  task-level bootstrap estimates.}
  \label{tab:oracle}
  \centering
  \small
  \resizebox{\textwidth}{!}{%
  \begin{tabular}{llrrll}
    \toprule
    Cohort & Model & Tasks & Admissible & Matches panel (\%) & Panel-confirmed suite (\%) \\
    \midrule
    Development & Qwen2.5 & 142 & 138 & 44.16 [39.65, 48.80] & 4.77 [3.36, 6.34] \\
    Development & Qwen3.5 & 142 & 137 & 58.90 [54.82, 62.88] & 11.50 [8.92, 14.16] \\
    External & Qwen2.5 & 114 & 112 & 27.79 [22.93, 32.89] & 2.53 [1.27, 4.09] \\
    External & Qwen3.5 & 114 & 93 & 50.12 [43.81, 56.43] & 5.07 [3.31, 7.02] \\
    \bottomrule
  \end{tabular}}
\end{table}

The \qthreefive condition is consistently better, but its error rate remains
too high for use as a reward oracle.  This is replication under a different
fixed model, not evidence of a scaling law.

\subsection{RQ2: One reference creates large apparent gains}

\begin{figure}[t]
  \centering
  \includegraphics[width=\textwidth]{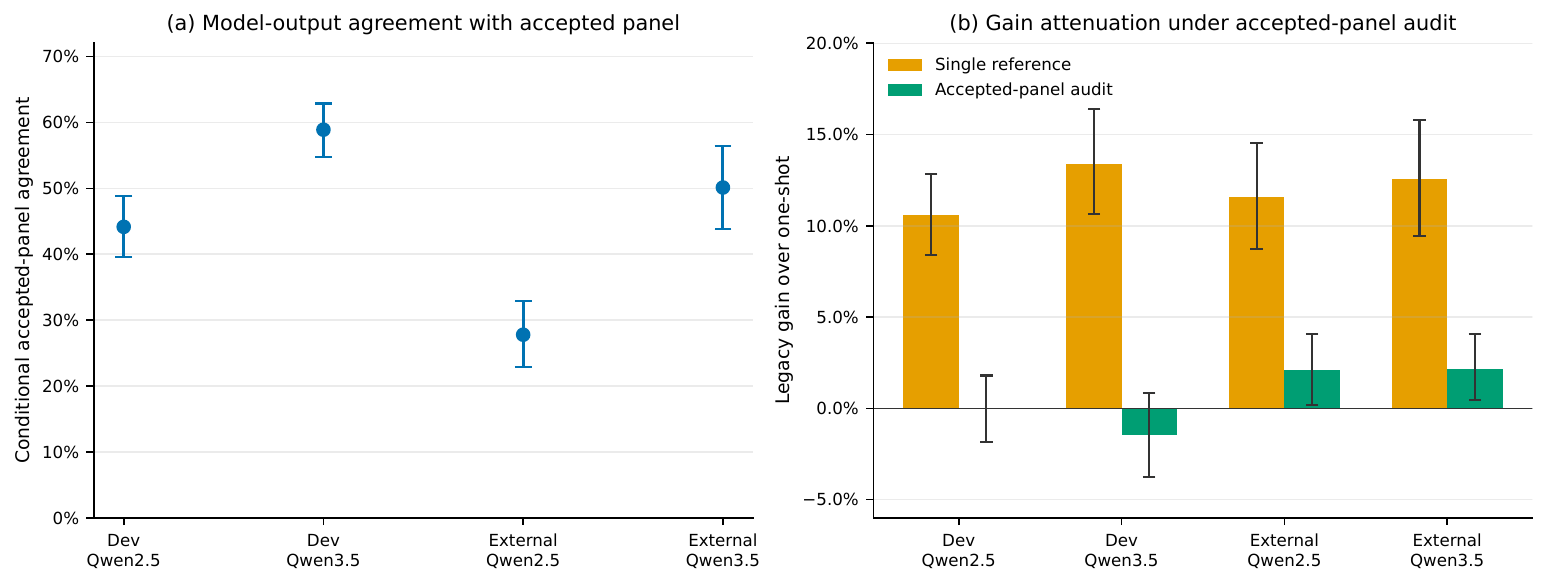}
  \caption{Oracle quality and apparent-versus-audited evolution gains.  Error
  bars are 95\% task-bootstrap intervals; cohorts are not pooled.}
  \Description{Two panels compare oracle reliability and the change in the
  estimated gain of evolution when a single reference is replaced by an audit
  over all accepted implementations. In every model and cohort, the audited gain
  is much smaller than the apparent gain.}
  \label{fig:oracle-feedback}
\end{figure}

Figure~\ref{fig:oracle-feedback} contrasts the nominal single-reference gain of
\methodlegacy over \methodone with the gain remaining after accepted-panel audit.
On development, the apparent gains are 10.58 and 13.38\pp; the same frozen
archives yield $-0.02$ and $-1.47$\pp after audit.  Externally, the apparent
11.58 and 12.57\pp gains shrink to 2.12 and 2.18\pp.  Consequently, estimated
gain inflation is 9.46--14.85\pp across all four conditions (Table
\ref{tab:gains}), with every 95\% interval strictly above zero.

\begin{table}[t]
  \caption{Evolution minus one-shot held-out constrained kill before and after
  auditing the frozen archive. All values are percentage points.}
  \label{tab:gains}
  \centering
  \small
  \resizebox{\textwidth}{!}{%
  \begin{tabular}{llrrr}
    \toprule
    Cohort & Model & Apparent [95\% CI] & Audited [95\% CI] & Inflation [95\% CI] \\
    \midrule
    Development & Qwen2.5 & 10.58 [8.41, 12.85] & $-0.02$ [$-1.82$, 1.80] & 10.60 [7.98, 13.47] \\
    Development & Qwen3.5 & 13.38 [10.67, 16.39] & $-1.47$ [$-3.74$, 0.85] & 14.85 [11.29, 18.60] \\
    External & Qwen2.5 & 11.58 [8.71, 14.55] & 2.12 [0.18, 4.09] & 9.46 [6.46, 12.76] \\
    External & Qwen3.5 & 12.57 [9.46, 15.82] & 2.18 [0.44, 4.08] & 10.39 [7.24, 13.77] \\
    \bottomrule
  \end{tabular}}
\end{table}

Adding only a second accepted implementation reduces apparent gains to
3.60--4.32\pp; three reduce them to 1.11--2.38\pp.  The full audit curves in the
supplement show that most inflation is detectable with one additional program.

\subsection{RQ3: Auditing reverses the method ranking}

Table~\ref{tab:methods-results} reports accepted-panel-audited held-out constrained
kill.  On development, \methodlegacy is indistinguishable from or slightly below
\methodone.  On the external cohort it is 2.12--2.18\pp higher, but this modest
gain does not survive the equal-candidate-budget comparison.  \methodmatched
exceeds \methodlegacy by 18.83 and 9.72\pp on development and by 10.91 and
6.01\pp externally; all four paired intervals lie strictly below zero when
expressed as \methodlegacy minus \methodmatched.

\begin{table}[t]
  \caption{Accepted-panel-audited held-out constrained kill (\%) and paired
  differences (percentage points). ``Public'' mutates public examples and is a
  diagnostic rather than a call-matched model baseline.}
  \label{tab:methods-results}
  \centering
  \small
  \resizebox{\textwidth}{!}{%
  \begin{tabular}{llrrrrll}
    \toprule
    Cohort & Model & One-shot & Evolve & Matched & Public & Evolve--One & Evolve--Matched \\
    \midrule
    Development & Qwen2.5 & 20.43 & 20.41 & 39.24 & 42.39 & $-0.02$ [$-1.78$,1.76] & $-18.83$ [$-21.95$,$-15.78$] \\
    Development & Qwen3.5 & 40.57 & 39.10 & 48.82 & 42.34 & $-1.47$ [$-3.79$,0.83] & $-9.72$ [$-12.44$,$-7.12$] \\
    External & Qwen2.5 & 10.58 & 12.70 & 23.61 & 27.63 & 2.12 [0.18,4.08] & $-10.91$ [$-14.91$,$-7.11$] \\
    External & Qwen3.5 & 12.91 & 15.09 & 21.10 & 27.63 & 2.18 [0.45,4.08] & $-6.01$ [$-9.15$,$-3.11$] \\
    \bottomrule
  \end{tabular}}
\end{table}

Public-input mutation shows that in-domain seeds have potential, but its
information source and call cost differ.  The conclusion concerns search
quality at nine candidate slots, not token or wall-clock efficiency.

The external two-point \methodlegacy gain over \methodone is metric-sensitive.
Held-out task detection changes by $-1.02$\pp for Qwen2.5 and $+0.15$\pp for
Qwen3.5, with both intervals crossing zero. Audited task coverage falls by 3.95 and
2.34\pp, respectively. The kill-rate gain does not broaden audited task
coverage, i.e., the fraction of tasks retaining a panel-confirmed test.

\subsection{RQ4: Unsafe mutations concentrate audit-disconfirmed credit}

Table~\ref{tab:fault-families} decomposes the
2,836 development faults.  Public and official exact-output anchors already
expose 41.61\% and 27.79\%, respectively.  Only 12.34\% fall into the four safe
local-mutation families combined.  A similarly sized 11.28\% family is exposed
only through public mutations that are audit-disconfirmed by the accepted panel.

\begin{table}[t]
  \caption{Selected fault-family results. Audit-disconfirmed share is the
  fraction of single-reference evolution detections rejected by the
  accepted-program panel; it is not a semantic false-positive rate.}
  \label{tab:fault-families}
  \centering
  \small
  \resizebox{\textwidth}{!}{%
  \begin{tabular}{lrrr}
    \toprule
    Trigger phenotype & Faults (\%) & Qwen2.5 audit-disconfirmed (\%) & Qwen3.5 audit-disconfirmed (\%) \\
    \midrule
    Public exposed & 1,180 (41.61) & 25.44 & 17.35 \\
    Official-oracle exposed & 788 (27.79) & 43.26 & 39.06 \\
    Safe local mutation families & 350 (12.34) & --- & --- \\
    Unsafe-mutation only & 320 (11.28) & 92.45 & 84.52 \\
    Hard residual & 198 (6.98) & 56.00 & 44.12 \\
    \bottomrule
  \end{tabular}}
\end{table}

The stronger condition reduces but does not eliminate disconfirmation.  No
family has a 95\% lower bound above zero for \methodlegacy versus
\methodmatched in both models; the largest exploratory exception also crosses
zero.  No stable subgroup reverses the ranking.

The robustness appendix repeats the paired comparisons for each seed and for
held-out kill, task detection, and audited task coverage. All 12 model--cohort--seed
gain-inflation intervals have positive lower bounds, and all 12 audited
\methodlegacy-minus-\methodmatched intervals have negative upper bounds.  The
same latter result holds for all 12 model--cohort--metric cells.

\subsection{RQ5: Semantic review supports the dominant mechanism but preserves exceptions}

The two blinded doctoral reviewers assign the same five-class label to 275 of
300 sampled inputs (91.67\%; inter-rater Cohen's $\kappa=.823$).  Collapsing the
labels to valid versus other, they agree on 293 inputs (97.67\%;
$\kappa=.825$).  Thus, 25 cases differ at the five-class level, but only seven
cross the valid--invalid boundary; the remaining disagreements concern the two
invalidity subtypes.

Both reviewers classify 275 inputs as invalid and 18 as valid; seven cross the
valid--invalid boundary.  For jointly valid cases, derived outputs support the
primary output in 13, an auditor output in one, and neither unambiguously in
four.  Relaxed limits restore panel agreement in three cases and fail to
reproduce one primary output.  These exceptions preclude treating panel
disagreement as semantic truth.

Table~\ref{tab:semantic-audit} reports population estimates recovered using the
locked sampling design.  Overall, 94.41\% (95\% stratified-bootstrap CI
[91.85, 96.51]) are jointly judged invalid, while 3.60\% [2.23, 5.14] are
jointly judged valid.  Weighting sampled inputs by their held-out kill links
gives the same qualitative result: 94.91\% [91.12, 97.62] invalid and 2.77\%
[1.34, 4.54] valid.  Development and external estimates are similar.  The
matched-resampling subgroup has a substantially larger jointly valid estimate
than native or public mutation, although this post-hoc subgroup comparison is
descriptive and not a method-selection result.

\begin{table}[t]
  \caption{Blinded human validity judgments for sampled
  audit-disconfirmed inputs. Rates are design-weighted; only the overall row
  shows pre-specified stratified-bootstrap 95\% CIs.}
  \label{tab:semantic-audit}
  \centering
  \small
  \begin{tabular}{lrrr}
    \toprule
    Group & Sample $n$ & Both invalid (\%) & Both valid (\%) \\
    \midrule
    Overall & 300 & 94.41 [91.85, 96.51] & 3.60 [2.23, 5.14] \\
    Development & 150 & 94.28 & 4.89 \\
    External & 150 & 94.61 & 1.57 \\
    Native evolve & 196 & 96.25 & 2.42 \\
    Matched resampling & 50 & 66.18 & 27.91 \\
    Public evolve & 50 & 98.14 & 0.00 \\
    One shot & 4 & 75.00 & 25.00 \\
    \bottomrule
  \end{tabular}
\end{table}

The 67-case follow-up packet contains all inter-rater disagreements and jointly
valid cases plus consensus quality controls.  It is a targeted adjudication set,
not an error count; the reported estimates do not force consensus or include a
third-reviewer resolution.

\subsection{RQ6: Available evidence does not isolate a robust aligned-feedback benefit}

Table~\ref{tab:feedback-placebo} separates the observed effect of entering a
three-round interaction from the incremental effect of behavior-aligned
execution feedback.  Against plain equal-candidate-slot resampling, real feedback
decreases audited held-out kill by 6.11\pp in the Qwen2.5 condition, but
increases it by 2.73\pp in the Qwen3.5 condition.  Read alone, E14 would suggest
a model-dependent benefit from feedback.

The placebo reproduces that direction and nearly the same magnitude: it changes
kill by $-6.24$ and $+3.23$\pp.  Real feedback exceeds placebo by only
$+0.13$\pp (95\% CI [$-1.56$, $+1.80$]) for Qwen2.5 and is $-0.50$\pp
[$-2.13$, $+1.08$] lower for Qwen3.5.  The cross-model interaction in this
real-minus-placebo contrast is $-0.63$\pp [$-2.92$, $+1.70$].  None of these
aligned-feedback intervals excludes zero.

\begin{table}[t]
  \caption{External accepted-panel-audited held-out constrained kill.  Levels
  are percentages; contrasts and 95\% paired task-bootstrap CIs are percentage
  points.  The placebo analysis is post hoc.}
  \label{tab:feedback-placebo}
  \centering
  \small
  \resizebox{\textwidth}{!}{%
  \begin{tabular}{lrrrll}
    \toprule
    Model & Plain & Placebo & Real & Placebo--Plain [95\% CI] & Real--Placebo [95\% CI] \\
    \midrule
    Qwen2.5 & 23.63 & 17.38 & 17.51 & $-6.24$ [$-8.64$,$-3.93$] & $+0.13$ [$-1.56$,1.80] \\
    Qwen3.5 & 21.18 & 24.42 & 23.92 & $+3.23$ [0.85,5.76] & $-0.50$ [$-2.13$,1.08] \\
    \bottomrule
  \end{tabular}}
\end{table}

The placebo is intentionally conservative: it retains detected-set size and
the largest per-input kill-count density while randomizing the input--fault
mapping.  Consequently, real minus placebo isolates the observed value of
fine-grained attribution, whereas placebo minus plain combines prompt/history
scaffolding with coarse progress signals.  The result does not show that all
execution feedback is useless, nor do zero-crossing intervals prove
equivalence. This is the E15 post-hoc diagnostic; E16-B supplies an
outcome-blind but qualification-amended held-out boundary check.

Table~\ref{tab:e16-feedback} reports E16-B. Placebo--Plain is $-11.39$\pp for
Qwen2.5 and $+4.47$\pp for Qwen3.5. Real--Placebo is much smaller: $+1.99$\pp
[$+0.08$,$+3.88$] and $+0.28$\pp [$-1.41$,$+2.03$], respectively. The first
interval fails the frozen superiority threshold, and the second misses the
equivalence boundary by 0.03\pp. Both model-specific decisions are inconclusive;
the amended cohort therefore neither establishes equivalence nor a practically
important advantage. The cross-model contrast is also unresolved
($-1.70$\pp, 95\% CI [$-4.09$,$+0.67$]).

\begin{table}[t]
  \caption{E16-B accepted-panel-audited held-out constrained kill.  Levels are
  percentages; contrasts and 95\% paired task-bootstrap CIs are percentage
  points.  ``Inc.'' denotes the frozen inconclusive decision.}
  \label{tab:e16-feedback}
  \centering
  \small
  \resizebox{\textwidth}{!}{%
  \begin{tabular}{lrrrlll}
    \toprule
    Model & Plain & Placebo & Real & Placebo--Plain [95\% CI] & Real--Placebo [95\% CI] & Decision \\
    \midrule
    Qwen2.5 & 42.42 & 31.03 & 33.01 & $-11.39$ [$-14.44$,$-8.48$] & $+1.99$ [$+0.08$,$+3.88$] & Inc. \\
    Qwen3.5 & 49.91 & 54.37 & 54.66 & $+4.47$ [$+2.28$,$+6.74$] & $+0.28$ [$-1.41$,$+2.03$] & Inc. \\
    \bottomrule
  \end{tabular}}
\end{table}

Source-interaction and secondary-outcome intervals cross zero (supplemental
material). Per direction-round, Qwen2.5 yields 0.837 confirmed Real candidates
and 0.794 Placebo candidates. Qwen3.5 is nearly matched at 1.874 and 1.882.
Equal slots therefore do not guarantee equal usable opportunity.

Table~\ref{tab:e16-yield-matched} reports the post-hoc replay. Reference-valid
matching changes the two contrasts to $-0.30$ and $+0.78$\pp;
safety-stratified matching gives $-0.35$ and $+0.13$\pp. All intervals in the
table cross zero. Because truncation is post-treatment and later rounds retain
original histories, this shows non-robustness rather than a direct effect. In
combination, E15 and E16-B narrow the plausible claim but do not constitute a
clean confirmatory test of feedback efficacy or ineffectiveness.

\begin{table}[t]
  \caption{Post-hoc E16-B candidate-yield-matched replay.  Entries are
  accepted-panel-audited Real--Placebo held-out kill in percentage points with
  95\% task-bootstrap CIs.}
  \label{tab:e16-yield-matched}
  \centering
  \small
  \begin{tabular}{lll}
    \toprule
    Replay & Qwen2.5 & Qwen3.5 \\
    \midrule
    Frozen original & $+1.99$ [$+0.08$,$+3.88$] & $+0.28$ [$-1.41$,$+2.03$] \\
    Reference-valid matched & $-0.30$ [$-0.90$,$+0.30$] & $+0.78$ [$-0.62$,$+2.19$] \\
    Safety-stratified matched & $-0.35$ [$-0.91$,$+0.20$] & $+0.13$ [$-1.03$,$+1.23$] \\
    \bottomrule
  \end{tabular}
\end{table}

\section{Discussion}
\label{sec:discussion}

\subsection{Execution is conditional verification}

Execution verifies a test only relative to a valid input domain and an adequate
oracle. One accepted program supplies neither. It can produce an output outside
the specification, turning disagreement with historical programs into apparent
fault detection. Repeated optimization can then convert this artifact into
training data or reward. Better output generation does not remove the risk:
larger search can still find inputs on which the reference is permissive or
idiosyncratic.

The semantic audit triangulates this mechanism. Two blinded software engineering
doctoral reviewers jointly classify 94.41\% of sampled panel-disconfirmed inputs
as invalid, but 3.60\% as valid and one valid case supports an auditor-side
output. Shared interpretation bias and reviewer error remain possible. Panel
execution is therefore a conservative consistency filter, not semantic ground
truth.

\subsection{An iterative loop is not a feedback effect}

Real feedback appears to hurt Qwen2.5 and help Qwen3.5 relative to resampling,
but the density-matched placebo reproduces both directions. On the amended
held-out cohort, the Real--Placebo contrast remains inconclusive under the
pre-specified $\pm2$\pp rule, and
the Qwen2.5 contrast disappears after equalizing observed valid yield. The
stronger condition follows the richer interface more reliably; the weaker one
loses candidates through format failures.

This does not show that execution feedback is universally useless. The placebo
retains aggregate progress and removes only fine-grained input--fault alignment.
The supported conclusion is narrower: scaffolding or coarse progress explains
the dominant model-dependent shift observed on the exposed external cohort,
while the residual value of aligned attribution remains unresolved and
opportunity-sensitive. E15 is post hoc and E16-B changes qualification after
seeing task supply; together they provide triangulated diagnostic evidence,
not confirmation that aligned feedback has zero effect.

\subsection{Evaluation safeguards}

Future feedback-driven test studies should apply the following controls:

\begin{enumerate}
  \item \textbf{Freeze, then audit.} Differentially execute selected tests on
  at least two independent accepted programs and remove disconfirmed entries
  without replacement. Report an audit-aware method separately if desired.

  \item \textbf{Match search and feedback interfaces.} Compare evolution with
  independent generation at equal candidate slots, report call and execution
  costs, and add a placebo that preserves history and coarse feedback while
  breaking the claimed behavior alignment.

  \item \textbf{Separate selection from evaluation.} Cross-fit real faulty
  programs, use tasks as the unit of inference, and report kill, task detection,
  audited task coverage, parse yield, and oracle accuracy together; distinguish
  this task-level validity yield from line or branch coverage.

  \item \textbf{Adjudicate disagreements.} Resolve panel disagreement against
  the natural-language contract using executable constraints, bounded
  brute-force oracles, resource-limit replay, or independent human review;
  never equate disagreement with a false test by definition.
\end{enumerate}

These controls complement industrial pipelines that admit tests only after
build, pass, reliability, and coverage checks~\cite{alshahwan2024testgen}. In
natural-language-only algorithmic tasks, however, no trusted project test suite
serves as a final oracle.

\subsection{Implications for self-evolving tests}

Feedback-driven generation remains promising when search stays inside a
defensible contract. One direction is relational evolution: transform a public
example under an executable precondition and verify a metamorphic output
relation~\cite{segura2016metamorphic}. Historical execution can then choose or
rank supported relations instead of overriding hard validity failures. This is
future work, not a result of the present study.

The sampled disconfirmed population also suggests optimizing admissible
hard-case yield rather than raw disagreement. Jointly valid estimates are
27.91\% for matched resampling, 2.42\% for native evolution, and 0\% for public
evolution, although these post-hoc subgroups are descriptive. A prospective
study should lock this objective and validate it with executable contracts
before any parameter update.

\subsection{Cost and reproducibility}

The four primary model--cohort conditions used 4,608 model calls and 555,610
unique program--input executions. E14/E15 added 5,472 calls and 190,695
executions; E16-B added 9,108 calls and 197,642 executions. Two accepted
programs capture most of the ranking change, so audits can remain an offline
evaluation standard rather than a deployment-time cost. Hash-locked manifests
and task-level rows allow alternative analyses without rerunning untrusted
programs.

Persistent model updates are meaningful only if their gains survive independent
measurement. Our contribution is therefore an audit and attribution protocol,
not a new optimizer: establish a deployable oracle contract and isolate the
information in feedback before comparing SFT, preference optimization, or
reinforcement-learning updates.

\section{Related Work}
\label{sec:related}

\subsection{LLM-based test generation}

LLMs generate tests for ranking code, improving structural coverage, and
repairing existing suites. CodeT ranks generated programs by agreement with
generated tests~\cite{chen2023codet}. CodaMOSA invokes a code model when
search-based generation reaches a coverage plateau~\cite{lemieux2023codamosa},
and CoverUp iteratively feeds Python coverage gaps back to an LLM
~\cite{pizzorno2025coverup}. TestGen-LLM improves existing tests subject to
build, pass, reliability, and coverage filters~\cite{alshahwan2024testgen};
IntUT makes test intentions explicit for industrial Java projects
~\cite{nan2025intut}. These systems primarily study executable repository tests
and structural adequacy; broader evaluations find strong sensitivity to model,
prompt, and project choices~\cite{yang2024evaluation}. Their results also show
why validity cannot be reduced to coverage: data filtering materially affects
both coverage and bug detection~\cite{zhang2025less}.

Algorithmic benchmarks instead make solution discrimination explicit.
TestCase-Eval pairs 500 tasks with over 100,000 human submissions to measure
fault coverage and targeted exposure~\cite{yang2025testcaseeval}; we use a
conservatively filtered Python exact-output subset. CodeContests-O iteratively
refines tests using execution on correct and incorrect solutions and optimizes
fidelity and discriminability~\cite{cai2026codecontestso}. Our question is not
whether an LLM can produce executable or discriminating tests, but whether a
reported improvement survives an oracle audit conducted only after selection
and a comparison with equal search opportunity.

\subsection{Iterative feedback and self-evolution}

Test-time refinement can improve outputs without weight updates. Self-Refine
uses a model as generator, critic, and refiner~\cite{madaan2023selfrefine};
Reflexion stores verbal feedback from environmental trials
~\cite{shinn2023reflexion}; and Self-Debugging conditions code revision on
execution and explanation~\cite{chen2024selfdebug}. Conversely, controlled
reasoning experiments show that intrinsic self-correction can stagnate or
degrade without external feedback~\cite{huang2024selfcorrect}. These findings
motivate separating the effect of another interaction round from the effect of
grounded feedback content.

Persistent adaptation raises the same attribution problem. Self-adapting
language models generate data and update directives~\cite{zweiger2025seal};
surveys classify self-evolving agents by the component and timing of adaptation
~\cite{gao2025selfevolving}. In code, CURE co-evolves coder and tester without
ground-truth solutions~\cite{wang2025cure}, and UTRL adversarially trains unit-
test and code generators~\cite{lee2026utrl}. Such methods make generated tests
part of a learning signal. Our study examines two prerequisites for crediting
that signal: an independently defensible oracle and a control that preserves
interaction structure while breaking the claimed behavior alignment.

\subsection{Test-oracle validity and adequacy}

The oracle problem asks how to decide whether observed behavior is correct
~\cite{barr2015oracle}. Recent LLM studies expose this as a distinct bottleneck.
Doc2OracLL studies how API documentation affects generated Java test oracles
~\cite{hossain2025doc2oracll}; Konstantinou et al. ask whether generated oracles
encode actual rather than expected behavior~\cite{konstantinou2024actual}; and
Molinelli et al. evaluate oracle usefulness on projects created after likely
training cutoffs~\cite{molinelli2025oracles}. These works evaluate assertions or
expected behavior. We study a complementary failure: an execution-derived
oracle can be internally consistent yet semantically unsafe because its input
lies outside the natural-language contract.

Differential implementations, formal contracts, and metamorphic relations can
reduce dependence on one absolute oracle~\cite{segura2016metamorphic}. Mutation
testing measures whether suites distinguish seeded faulty variants
~\cite{jia2011mutation}; we instead cross-fit real historical faults so that
selection and measurement use disjoint fault folds. Agreement among accepted
implementations remains a benchmark-specific empirical oracle, not proof of
specification conformance, which motivates the separate human semantic audit.

\begin{table}[t]
  \caption{Positioning against representative adjacent work by objective and
  methodological emphasis.}
  \label{tab:related-positioning}
  \centering
  \small
  \begin{tabularx}{\textwidth}{lYYY}
    \toprule
    Work & Setting & Primary signal or objective & Main methodological emphasis \\
    \midrule
    CodaMOSA / CoverUp & Repository tests & Structural coverage & Coverage-guided generation \\
    Doc2OracLL & Java tests & Documentation-grounded assertions & Oracle generation quality \\
    CodeContests-O & Algorithmic tasks & Fidelity and discriminability & Iterative test refinement \\
    CURE / UTRL & Code-model training & Coder--tester reward & Persistent co-evolution \\
    This work & Algorithmic tasks & Cross-fit real-fault kill after audit & Post-selection audit, equal-slot control, and feedback placebo \\
    \bottomrule
  \end{tabularx}
\end{table}

Accordingly, our contribution is neither a new generator nor another benchmark.
It connects three controls that adjacent work usually studies separately:
freeze-then oracle audit, candidate-budget matching, and a matched placebo for
the feedback interface. The resulting protocol tests whether
apparent self-evolution reflects better tests, more search, or exploitation of
the evaluator.

\section{Threats to Validity}
\label{sec:threats}

\paragraph{Construct validity.}
Accepted-program agreement is stronger than one-reference execution but does
not prove that an input conforms to natural language. Accepted programs can
contain or share defects, and legal outputs may have multiple textual forms.
We restrict the study to exact-output tasks and normalize whitespace, but do not
generalize to special checkers or floating-point tolerance. Trigger phenotypes
are execution-grounded categories, not source-level causes. The human semantic
audit is also not definitive ground truth: the two doctoral reviewers may share
training and interpretation biases, and neither used the allowed ambiguity
labels. The 67-case packet therefore remains for consensus or third-reviewer
specification adjudication.

The placebo preserves iterative structure, history size, detected-set
cardinality, and leading kill-count density. It does not remove all coarse
feedback, nor exactly match prompt tokens. Real--Placebo consequently estimates
fine-grained input--fault attribution only, not the value of every execution
signal.

\paragraph{Internal validity.}
Development tasks were exposed and the family analysis is post hoc. The locked
external study fixes tasks, models, budgets, folds, hypotheses, and analyses
before fresh generation, and archives are selected before auditor execution.
Recorded infrastructure and provider amendments preserve generated responses
but can still affect execution or judgment behavior. E14 uses an already
exposed cohort; E15 was designed after E14 and remains exploratory.

E16-A froze before opening its 200 held-out tasks but failed its 20-fault gate.
E16-B changed the fault panel to 18 after observing qualification supply, before
model outcomes. It is outcome-blind but qualification-amended. Included tasks
have more exact-compatible accepted programs than excluded tasks, and no
outcomes exist for excluded tasks, so profile analyses cannot eliminate
selection bias. Consequently, E16-B's frozen analysis rules do not restore the
evidential status of the failed E16-A design, and RQ6 is not presented as a
clean confirmatory experiment.

\paragraph{Statistical conclusion validity.}
Tests and executions are dependent within task; we average seeds within task and
bootstrap paired tasks. Cohorts are not pooled, and exploratory family tests are
not multiplicity-corrected. Intervals describe the observed task populations,
not arbitrary domains or models. E15 lacked a smallest effect of interest, so
zero-crossing intervals do not prove equivalence. E16-B specified $\pm2$\pp but
neither model satisfies the frozen equivalence or superiority rule. Source
interactions cross zero. Yield-matched replay conditions on a post-treatment
mediator and cannot regenerate later rounds, so it demonstrates non-robustness,
not a causal decomposition.

\paragraph{External validity.}
All tasks are algorithmic, all executed submissions are Python-family, and
TestCase-Eval is a new source rather than a new application domain. Both model
conditions are from the Qwen family and differ in checkpoint provenance and
warm start, so they neither establish cross-family generality nor identify a
scale effect. A size-matched code-instruction model from an independent family
is the highest-priority replication; repository tests, other languages, richer
checkers, and production systems also require study.

\paragraph{Execution and reproducibility.}
External executions use strict namespace and resource-limit attestations. The
older development runner lacks equivalent per-row isolation evidence, and
timeouts or environmental differences may still alter classifications. Hashes
link model shards, manifests, sources, generations, and task-level summaries.
The audit trail exposes the failed E16-A gate, E16-B amendment, and subsequent
runtime amendments, improving transparency without restoring pristine
prospective status.

\section{Conclusion}
\label{sec:conclusion}

One accepted implementation inflates apparent test-evolution gains by
9.46--14.85\pp. After independent audit, equal-slot resampling is
6.01--18.83\pp better than mutation-based evolution. A genuine three-round
loop also does not establish a fine-grained feedback benefit: its
model-dependent shift is largely reproduced by a density-matched placebo, and
qualification-amended held-out Real--Placebo contrasts meet neither the
pre-specified equivalence nor superiority rule. Equalizing observed valid yield
removes the positive Qwen2.5 contrast. These RQ6 analyses are diagnostic rather
than pristine confirmation. Blinded semantic judgments support invalid inputs
as the dominant mechanism while preserving a nonzero set of valid exceptions.

Execution is useful feedback only inside a valid measurement contract. Future
self-evolving test generators should freeze selection before multi-program
audit, cross-fit real faults, match search budgets, report coverage and
abstention, and compare aligned feedback with an interaction- and
density-matched placebo. These controls distinguish learning to test from
learning to exploit a verifier or merely follow a richer interface.

\section*{Data Availability}
\label{sec:data-availability}

An anonymized replication package containing protocol locks, task-level derived
metrics, analysis code, figure builders, and integrity instructions is available
at \url{https://anonymous.4open.science/r/llm-test-evolution-audit-8CF5/}.
Restricted model weights and source datasets are represented by checksums and
deterministic preparation instructions.

\bibliographystyle{ACM-Reference-Format}
\bibliography{references}

@article{barr2015oracle,
  author  = {Earl T. Barr and Mark Harman and Phil McMinn and Muzammil Shahbaz and Shin Yoo},
  title   = {The Oracle Problem in Software Testing: A Survey},
  journal = {IEEE Transactions on Software Engineering},
  volume  = {41},
  number  = {5},
  pages   = {507--525},
  year    = {2015},
  doi     = {10.1109/TSE.2014.2372785},
  url     = {https://doi.org/10.1109/TSE.2014.2372785}
}

@article{jia2011mutation,
  author  = {Yue Jia and Mark Harman},
  title   = {An Analysis and Survey of the Development of Mutation Testing},
  journal = {IEEE Transactions on Software Engineering},
  volume  = {37},
  number  = {5},
  pages   = {649--678},
  year    = {2011},
  doi     = {10.1109/TSE.2010.62},
  url     = {https://doi.org/10.1109/TSE.2010.62}
}

@article{segura2016metamorphic,
  author  = {Sergio Segura and Gordon Fraser and Ana Bel{\'e}n S{\'a}nchez and Antonio Ruiz-Cort{\'e}s},
  title   = {A Survey on Metamorphic Testing},
  journal = {IEEE Transactions on Software Engineering},
  volume  = {42},
  number  = {9},
  pages   = {805--824},
  year    = {2016},
  doi     = {10.1109/TSE.2016.2532875},
  url     = {https://doi.org/10.1109/TSE.2016.2532875}
}

@inproceedings{chen2023codet,
  author    = {Bei Chen and Fengji Zhang and Anh Nguyen and Daoguang Zan and Zeqi Lin and Jian{-}Guang Lou and Weizhu Chen},
  title     = {{CodeT}: Code Generation with Generated Tests},
  booktitle = {International Conference on Learning Representations},
  year      = {2023},
  url       = {https://openreview.net/forum?id=ktrw68Cmu9c}
}

@inproceedings{yang2025testcaseeval,
  author    = {Zheyuan Yang and Zexi Kuang and Xue Xia and Yilun Zhao},
  title     = {Can {LLM}s Generate High-Quality Test Cases for Algorithm Problems? {TestCase-Eval}: A Systematic Evaluation of Fault Coverage and Exposure},
  booktitle = {Proceedings of the 63rd Annual Meeting of the Association for Computational Linguistics (Volume 2: Short Papers)},
  pages     = {1050--1063},
  year      = {2025},
  publisher = {Association for Computational Linguistics},
  doi       = {10.18653/v1/2025.acl-short.82},
  url       = {https://aclanthology.org/2025.acl-short.82/}
}

@inproceedings{alshahwan2024testgen,
  author    = {Nadia Alshahwan and Jubin Chheda and Anastasia Finogenova and Beliz Gokkaya and Mark Harman and Inna Harper and Alexandru Marginean and Shubho Sengupta and Eddy Wang},
  title     = {Automated Unit Test Improvement Using Large Language Models at Meta},
  booktitle = {Companion Proceedings of the 32nd ACM International Conference on the Foundations of Software Engineering},
  pages     = {185--196},
  year      = {2024},
  publisher = {Association for Computing Machinery},
  doi       = {10.1145/3663529.3663839},
  url       = {https://doi.org/10.1145/3663529.3663839}
}

@article{yang2024evaluation,
  author  = {Lin Yang and Chen Yang and Shutao Gao and Weijing Wang and Bo Wang and Qihao Zhu and Xiao Chu and Jianyi Zhou and Guangtai Liang and Qianxiang Wang and Junjie Chen},
  title   = {On the Evaluation of Large Language Models in Unit Test Generation},
  journal = {arXiv preprint arXiv:2406.18181},
  year    = {2024},
  doi     = {10.48550/arXiv.2406.18181},
  url     = {https://arxiv.org/abs/2406.18181}
}

@inproceedings{cai2026codecontestso,
  author    = {Jianfeng Cai and Jinhua Zhu and Ruopei Sun and Kangwen Zhao and Dongyun Xue and Mingxiao Feng and Wengang Zhou and Houqiang Li},
  title     = {{CodeContests-O}: Powering {LLM}s via Feedback-Driven Iterative Test Case Generation},
  booktitle = {Findings of the Association for Computational Linguistics: ACL 2026},
  pages     = {1054--1072},
  year      = {2026},
  publisher = {Association for Computational Linguistics},
  doi       = {10.18653/v1/2026.findings-acl.53},
  url       = {https://aclanthology.org/2026.findings-acl.53/}
}

@inproceedings{wang2025cure,
  author    = {Yinjie Wang and Ling Yang and Ye Tian and Ke Shen and Mengdi Wang},
  title     = {Co-Evolving {LLM} Coder and Unit Tester via Reinforcement Learning},
  booktitle = {Advances in Neural Information Processing Systems 38},
  year      = {2025},
  doi       = {10.52202/085713-4809},
  url       = {https://proceedings.neurips.cc/paper_files/paper/2025/hash/d38653cdaa8e992549e1e9e1621610d7-Abstract-Conference.html}
}

@inproceedings{lee2026utrl,
  author    = {Dongjun Lee and Changho Hwang and Kimin Lee},
  title     = {Learning to Generate Unit Test via Adversarial Reinforcement Learning},
  booktitle = {International Conference on Learning Representations},
  year      = {2026},
  doi       = {10.48550/arXiv.2508.21107},
  url       = {https://arxiv.org/abs/2508.21107}
}

@article{zweiger2025seal,
  author  = {Adam Zweiger and Jyothish Pari and Han Guo and Ekin Aky{\"u}rek and Yoon Kim and Pulkit Agrawal},
  title   = {Self-Adapting Language Models},
  journal = {arXiv preprint arXiv:2506.10943},
  year    = {2025},
  doi     = {10.48550/arXiv.2506.10943},
  url     = {https://arxiv.org/abs/2506.10943}
}

@article{gao2025selfevolving,
  author  = {Huan{-}ang Gao and Jiayi Geng and Wenyue Hua and Mengkang Hu and Xinzhe Juan and Hongzhang Liu and Shilong Liu and Jiahao Qiu and Xuan Qi and Yiran Wu and others},
  title   = {A Survey of Self-Evolving Agents: On Path to Artificial Super Intelligence},
  journal = {arXiv preprint arXiv:2507.21046},
  year    = {2025},
  doi     = {10.48550/arXiv.2507.21046},
  url     = {https://arxiv.org/abs/2507.21046}
}

@article{hui2024qwen25coder,
  author  = {Binyuan Hui and Jian Yang and Zeyu Cui and Jiaxi Yang and Dayiheng Liu and Lei Zhang and Tianyu Liu and Jiajun Zhang and Bowen Yu and Keming Lu and others},
  title   = {{Qwen2.5-Coder} Technical Report},
  journal = {arXiv preprint arXiv:2409.12186},
  year    = {2024},
  doi     = {10.48550/arXiv.2409.12186},
  url     = {https://arxiv.org/abs/2409.12186}
}

@misc{qwen2026qwen35,
  author       = {{Qwen Team}},
  title        = {{Qwen3.5-9B} Model Card},
  year         = {2026},
  howpublished = {Hugging Face model repository},
  url          = {https://huggingface.co/Qwen/Qwen3.5-9B},
  note         = {Accessed 2026-08-15}
}

@article{cohen1960coefficient,
  author  = {Jacob Cohen},
  title   = {A Coefficient of Agreement for Nominal Scales},
  journal = {Educational and Psychological Measurement},
  volume  = {20},
  number  = {1},
  pages   = {37--46},
  year    = {1960},
  doi     = {10.1177/001316446002000104},
  url     = {https://doi.org/10.1177/001316446002000104}
}

@inproceedings{lemieux2023codamosa,
  author    = {Caroline Lemieux and Jeevana Priya Inala and Shuvendu K. Lahiri and Siddhartha Sen},
  title     = {{CodaMOSA}: Escaping Coverage Plateaus in Test Generation with Pre-trained Large Language Models},
  booktitle = {Proceedings of the 45th International Conference on Software Engineering},
  pages     = {919--931},
  year      = {2023},
  publisher = {IEEE},
  doi       = {10.1109/ICSE48619.2023.00085},
  url       = {https://doi.org/10.1109/ICSE48619.2023.00085}
}

@inproceedings{madaan2023selfrefine,
  author    = {Aman Madaan and Niket Tandon and Prakhar Gupta and Skyler Hallinan and Luyu Gao and Sarah Wiegreffe and Uri Alon and Nouha Dziri and Shrimai Prabhumoye and Yiming Yang and Shashank Gupta and Bodhisattwa Prasad Majumder and Katherine Hermann and Sean Welleck and Amir Yazdanbakhsh and Peter Clark},
  title     = {Self-Refine: Iterative Refinement with Self-Feedback},
  booktitle = {Advances in Neural Information Processing Systems 36},
  pages     = {46534--46594},
  year      = {2023},
  doi       = {10.52202/075280-2019},
  url       = {https://proceedings.neurips.cc/paper_files/paper/2023/hash/91edff07232fb1b55a505a9e9f6c0ff3-Abstract-Conference.html}
}

@inproceedings{shinn2023reflexion,
  author    = {Noah Shinn and Federico Cassano and Ashwin Gopinath and Karthik Narasimhan and Shunyu Yao},
  title     = {Reflexion: Language Agents with Verbal Reinforcement Learning},
  booktitle = {Advances in Neural Information Processing Systems 36},
  pages     = {8634--8652},
  year      = {2023},
  doi       = {10.52202/075280-0377},
  url       = {https://proceedings.neurips.cc/paper_files/paper/2023/hash/1b44b878bb782e6954cd888628510e90-Abstract-Conference.html}
}

@inproceedings{chen2024selfdebug,
  author    = {Xinyun Chen and Maxwell Lin and Nathanael Sch{\"a}rli and Denny Zhou},
  title     = {Teaching Large Language Models to Self-Debug},
  booktitle = {The Twelfth International Conference on Learning Representations},
  year      = {2024},
  url       = {https://openreview.net/forum?id=KuPixIqPiq}
}

@inproceedings{huang2024selfcorrect,
  author    = {Jie Huang and Xinyun Chen and Swaroop Mishra and Huaixiu Steven Zheng and Adams Wei Yu and Xinying Song and Denny Zhou},
  title     = {Large Language Models Cannot Self-Correct Reasoning Yet},
  booktitle = {The Twelfth International Conference on Learning Representations},
  year      = {2024},
  url       = {https://openreview.net/forum?id=IkmD3fKBPQ}
}

@article{pizzorno2025coverup,
  author  = {Juan Altmayer Pizzorno and Emery D. Berger},
  title   = {{CoverUp}: Effective High Coverage Test Generation for Python},
  journal = {Proceedings of the ACM on Software Engineering},
  volume  = {2},
  number  = {FSE},
  pages   = {2897--2919},
  year    = {2025},
  doi     = {10.1145/3729398},
  url     = {https://doi.org/10.1145/3729398}
}

@article{hossain2025doc2oracll,
  author  = {Soneya Binta Hossain and Raygan Taylor and Matthew B. Dwyer},
  title   = {{Doc2OracLL}: Investigating the Impact of Documentation on {LLM}-Based Test Oracle Generation},
  journal = {Proceedings of the ACM on Software Engineering},
  volume  = {2},
  number  = {FSE},
  pages   = {1870--1891},
  year    = {2025},
  doi     = {10.1145/3729354},
  url     = {https://doi.org/10.1145/3729354}
}

@article{zhang2025less,
  author  = {Junwei Zhang and Xing Hu and Shan Gao and Xin Xia and David Lo and Shanping Li},
  title   = {Less Is More: On the Importance of Data Quality for Unit Test Generation},
  journal = {Proceedings of the ACM on Software Engineering},
  volume  = {2},
  number  = {FSE},
  pages   = {1293--1316},
  year    = {2025},
  doi     = {10.1145/3715778},
  url     = {https://doi.org/10.1145/3715778}
}

@inproceedings{nan2025intut,
  author    = {Zifan Nan and Zhaoqiang Guo and Kui Liu and Xin Xia},
  title     = {Test Intention Guided {LLM}-Based Unit Test Generation},
  booktitle = {Proceedings of the 47th IEEE/ACM International Conference on Software Engineering},
  pages     = {1026--1038},
  year      = {2025},
  doi       = {10.1109/ICSE55347.2025.00243},
  url       = {https://doi.org/10.1109/ICSE55347.2025.00243}
}

@article{konstantinou2024actual,
  author  = {Michael Konstantinou and Renzo Degiovanni and Mike Papadakis},
  title   = {Do {LLM}s Generate Test Oracles That Capture the Actual or the Expected Program Behaviour?},
  journal = {arXiv preprint arXiv:2410.21136},
  year    = {2024},
  doi     = {10.48550/arXiv.2410.21136},
  url     = {https://arxiv.org/abs/2410.21136}
}

@inproceedings{molinelli2025oracles,
  author    = {Davide Molinelli and Luca Di Grazia and Alberto Martin-Lopez and Michael D. Ernst and Mauro Pezz{\`e}},
  title     = {Do {LLM}s Generate Useful Test Oracles? An Empirical Study with an Unbiased Dataset},
  booktitle = {Proceedings of the 40th IEEE/ACM International Conference on Automated Software Engineering},
  pages     = {278--290},
  year      = {2025},
  doi       = {10.1109/ASE63991.2025.00031},
  url       = {https://doi.org/10.1109/ASE63991.2025.00031}
}

\clearpage
\appendix
\section{Robustness Analyses}
\label{app:robustness}

These analyses were specified after observing the primary results and are
therefore robustness checks, not new decision gates. They make no model
calls, execute no programs, and do not authorize method selection.

\subsection{Audit depth}

For depth $k\in\{1,2,3\}$, we enumerate every size-$k$ subset of available
accepted implementations within each task-repeat and average the result before
averaging repeat seeds within task.  ``Full'' uses all 3--10 development
implementations and all three external implementations.  Depth one exactly
reconstructs the saved single-reference analysis; Full exactly reconstructs the
saved accepted-panel analysis.

\begin{table}[h]
  \caption{\methodlegacy minus \methodone held-out constrained kill (percentage
  points) as audit depth increases.}
  \label{tab:audit-depth}
  \centering
  \small
  \begin{tabular}{llrrrr}
    \toprule
    Cohort & Model & 1 ref. & 2 refs. & 3 refs. & Full \\
    \midrule
    Development & Qwen2.5 & 10.58 & 4.32 & 2.38 & $-0.02$ \\
    Development & Qwen3.5 & 13.38 & 3.66 & 1.11 & $-1.47$ \\
    External & Qwen2.5 & 11.58 & 3.60 & 2.12 & 2.12 \\
    External & Qwen3.5 & 12.57 & 3.99 & 2.18 & 2.18 \\
    \bottomrule
  \end{tabular}
\end{table}

\begin{figure}[h]
  \centering
  \includegraphics[width=\textwidth]{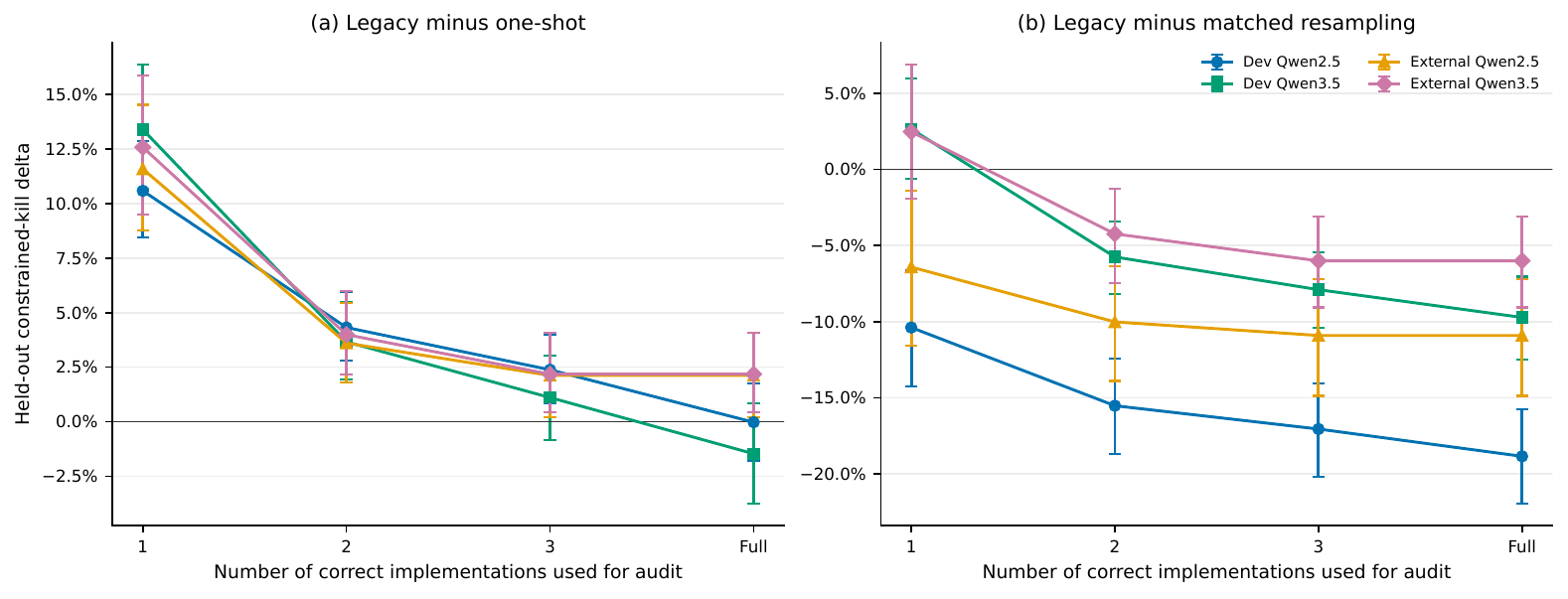}
  \caption{Method differences as independent oracle references are added.}
  \Description{Four audit-depth curves show that the estimated advantage of
  evolution over one shot decreases sharply as references are added, while its
  disadvantage to matched resampling grows.}
  \label{fig:audit-depth}
\end{figure}

Notably, the Qwen3.5 condition ranks \methodlegacy slightly above
\methodmatched with one reference on both cohorts, although the intervals cross
zero.  With only two references the differences become $-5.75$\pp (95\% CI
[$-8.17$,$-3.42$]) on development and $-4.24$\pp
[$-7.46$,$-1.26$] externally.  Method ranking therefore does not depend on
auditing with a large ensemble.

\subsection{Seed and metric sensitivity}

\begin{figure}[h]
  \centering
  \includegraphics[width=\textwidth]{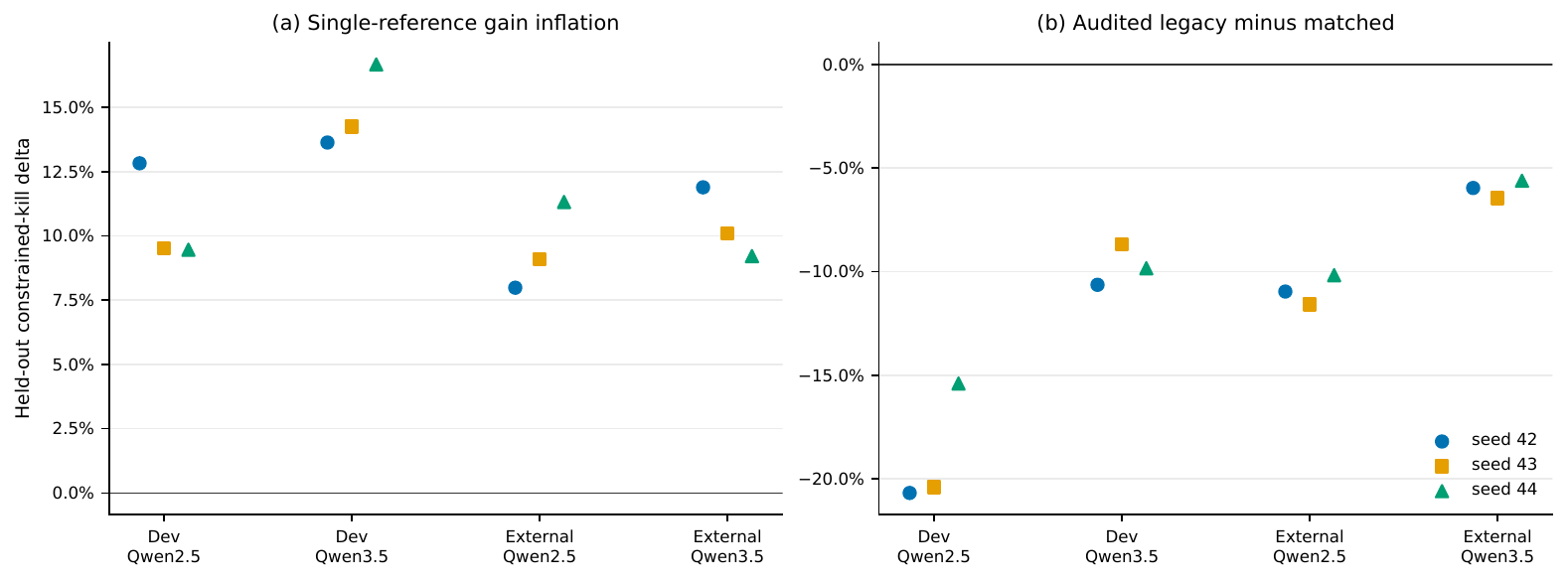}
  \caption{Seed-wise gain inflation and audited method differences.}
  \Description{Points and intervals for three seeds in four conditions show
  positive gain inflation and negative audited evolution-minus-matched effects
  throughout.}
  \label{fig:seed-stability}
\end{figure}

Figure~\ref{fig:seed-stability} reports seeds 42, 43, and 44 separately.  In all
12 model--cohort--seed cells, the lower confidence bound for gain inflation is
positive.  In all 12 cells, the upper bound for audited \methodlegacy minus
\methodmatched is negative.  Individual external audited gains over one shot are
small and often uncertain, whereas the matched comparison is directionally
stable.

Replacing held-out constrained kill with held-out task detection or archive task
coverage yields the same conclusion against matched resampling: all 12
model--cohort--metric upper confidence bounds are negative.  The external
evolution-minus-one-shot kill gains of 2.12/2.18\pp correspond to task-detection
changes of $-1.02$/$+0.15$\pp and coverage changes of $-3.95$/$-2.34$\pp.

\begin{figure}[h]
  \centering
  \includegraphics[width=\textwidth]{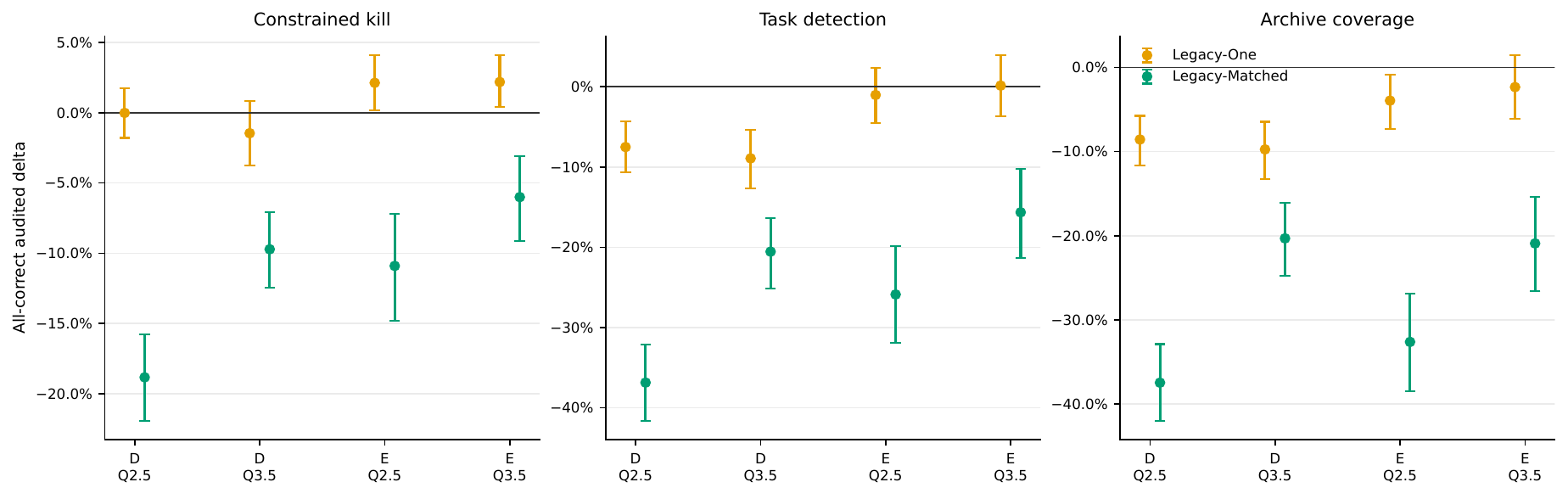}
  \caption{Accepted-panel-audited method differences under three outcomes.}
  \Description{Across constrained kill, task detection, and task coverage,
  evolution is consistently below matched resampling in all four conditions.}
  \label{fig:metric-sensitivity}
\end{figure}

\subsection{Cost accounting}

\begin{table}[h]
  \caption{Observed historical costs. Evaluation hours are summed across
  executions and are not wall-clock runtime.}
  \label{tab:cost}
  \centering
  \small
  \resizebox{\textwidth}{!}{%
  \begin{tabular}{llrrrrr}
    \toprule
    Cohort & Model & Calls & Prompt tok. & Completion tok. & Executions & Eval. h \\
    \midrule
    Development & Qwen2.5 & 1,278 & 796,311 & 208,781 & 168,017 & 11.65 \\
    Development & Qwen3.5 & 1,278 & 794,772 & 184,437 & 200,234 & 13.62 \\
    External & Qwen2.5 & 1,026 & 653,859 & 189,424 & 94,025 & 4.49 \\
    External & Qwen3.5 & 1,026 & 652,221 & 169,963 & 93,334 & 4.69 \\
    \bottomrule
  \end{tabular}}
\end{table}

The total is 4,608 calls and 555,610 unique program--input executions.  External
rows attest strict bubblewrap isolation; the older development rows record
limits and counts but not equivalent per-row isolation.  The robustness
reanalysis adds no generation or execution cost.

\section{Three-Round Feedback Diagnostic}
\label{app:feedback}

Table~\ref{tab:feedback-views} reports both oracle views for the E14/E15
branches.  The single-reference view makes the apparent Qwen3.5 gain larger and
the Qwen2.5 loss larger, but does not change the real-versus-placebo conclusion.
The audited cross-model interaction for real minus placebo is $-0.63$\pp (95\%
CI [$-2.92$, $+1.70$]).

\begin{table}[h]
  \caption{Held-out constrained kill (\%) for the external three-round study.}
  \label{tab:feedback-views}
  \centering
  \small
  \begin{tabular}{llrrr}
    \toprule
    Oracle view & Model & Plain & Placebo & Real \\
    \midrule
    Single reference & Qwen2.5 & 31.04 & 21.49 & 22.69 \\
    Single reference & Qwen3.5 & 24.99 & 30.77 & 31.15 \\
    Accepted-panel audited & Qwen2.5 & 23.63 & 17.38 & 17.51 \\
    Accepted-panel audited & Qwen3.5 & 21.18 & 24.42 & 23.92 \\
    \bottomrule
  \end{tabular}
\end{table}

Placebo parse rates in rounds one and two are 23.83\% and 26.02\% for Qwen2.5,
and 77.49\% and 83.33\% for Qwen3.5. Real-feedback rates are 25.00\% and
27.49\%, and 78.95\% and 82.46\%, respectively. This close tracking is a manipulation check:
the placebo retains the interaction burden that plausibly causes the Qwen2.5
degradation.  It does not by itself prove which part of the iterative interface
causes the Qwen3.5 improvement.

E14 and E15 each make 1,368 new calls per model.  Combined, they add 5,472
calls, 4,893,942 prompt tokens, 911,550 completion tokens, and 190,695 unique
program--input executions.  E14's procedure was prospectively locked on the
already exposed external cohort; E15 was designed after E14 and is post hoc.
Both bundles record strict bubblewrap isolation, source hashes, generated-row
hashes, and summary locks.

\section{Qualification-Amended Held-Out Feedback Follow-up}
\label{app:e16}

E16-A opened 100 IID and 100 OOD tasks only after freezing the qualification
rule, but stopped before model calls because the 20-fault rule retained only 36
OOD tasks.  The signed failure record is preserved.  E16-B was then frozen
before model outcomes, changed the fault panel to 18 with balanced 9/9 folds,
and retained 138 tasks (71 IID, 67 OOD).  One otherwise eligible task exceeded
the frozen 1,800-token prompt limit.  This is an outcome-blind qualification
amendment, not a pristine confirmation.

The complete qualification flow starts from 200 tasks and retains 138.  The 62
excluded tasks comprise 27 with accepted-program output disagreement alone, 18
with both disagreement and too few exact-compatible references, 4 with too few
references alone, 11 with too few faults alone, 1 with too few references and
oracle cases, and 1 over the prompt limit.  Thus, the nonexclusive reason counts
are 45 disagreements, 23 insufficient exact references, 11 insufficient faults,
1 insufficient oracle case, and 1 prompt overflow.

\begin{table}[h]
  \caption{E16-B qualification profiles.  Means describe observables available
  before model generation.}
  \label{tab:e16-qualification-profile}
  \centering
  \small
  \resizebox{\textwidth}{!}{%
  \begin{tabular}{lrrrrr}
    \toprule
    Group & Tasks & IID & OOD & Exact-compatible accepted & Fault candidates \\
    \midrule
    Included & 138 & 71 & 67 & 9.91 & 19.68 \\
    Excluded & 62 & 29 & 33 & 4.98 & 19.29 \\
    \bottomrule
  \end{tabular}}
\end{table}

\begin{table}[h]
  \caption{Fault-threshold supply after all non-fault qualification filters but
  before prompt-length filtering.  Newly admitted tasks below 18 faults were
  not prompt-tokenized in the frozen run.}
  \label{tab:e16-threshold-supply}
  \centering
  \small
  \begin{tabular}{rrrr}
    \toprule
    Minimum faults & IID & OOD & Total \\
    \midrule
    15 & 71 & 78 & 149 \\
    16 & 71 & 78 & 149 \\
    17 & 71 & 76 & 147 \\
    18 & 71 & 68 & 139 \\
    19 & 71 & 55 & 126 \\
    20 & 71 & 36 & 107 \\
    \bottomrule
  \end{tabular}
\end{table}

Changing the threshold therefore changes almost exclusively the OOD stratum.
On the evaluated nested subsets with at least 18, 19, and 20 available faults,
Qwen2.5 Real--Placebo is $+1.99$, $+1.79$, and $+1.59$\pp; Qwen3.5 is
$+0.28$, $+0.70$, and $+1.33$\pp.  The 19- and 20-fault intervals cross zero
for both models, and the point estimates do not share a monotone direction.
This is a post-hoc composition sensitivity, not evidence about unevaluated
tasks below 18 faults.

\begin{table}[h]
  \caption{E16-B held-out constrained kill (\%) under both oracle views.}
  \label{tab:e16-feedback-views}
  \centering
  \small
  \begin{tabular}{llrrr}
    \toprule
    Oracle view & Model & Plain & Placebo & Real \\
    \midrule
    Single reference & Qwen2.5 & 47.81 & 35.12 & 37.28 \\
    Single reference & Qwen3.5 & 51.92 & 57.58 & 57.50 \\
    Accepted-panel audited & Qwen2.5 & 42.42 & 31.03 & 33.01 \\
    Accepted-panel audited & Qwen3.5 & 49.91 & 54.37 & 54.66 \\
    \bottomrule
  \end{tabular}
\end{table}

Under one reference, Real--Placebo is $+2.16$\pp for Qwen2.5 and $-0.08$\pp
for Qwen3.5. Panel auditing changes these to $+1.99$ and $+0.28$\pp,
respectively; the main paper reports their confidence intervals. Neither lies
wholly inside the
pre-specified $\pm2$\pp equivalence region, and neither has the lower bound above the
$+2$\pp practical-superiority threshold. Both model-specific decisions are
therefore inconclusive within this qualification-amended cohort. E16-B uses
9,108 model calls, 7,725,394 prompt tokens, 1,483,998
completion tokens, and 197,642 unique program--input executions.

Table~\ref{tab:e16-process} reports the secondary process measures omitted from
the primary kill table.  Audit rejection is the ratio of selected
single-reference archive entries removed by the accepted panel.  Candidate
counts are per distinct model call; duplicate counts are candidates rejected
because the same normalized input appeared earlier in that branch.  The final
column divides audited held-out faults killed by total tokens and is descriptive
rather than an inferential efficacy estimand.  For arm-wise costing, the shared
round-zero call is charged once to each arm; actual experiment-wide totals above
count it only once.

\begin{table}[h]
  \caption{E16-B process, coverage, and descriptive efficiency diagnostics.}
  \label{tab:e16-process}
  \centering
  \scriptsize
  \resizebox{\textwidth}{!}{%
  \begin{tabular}{llrrrrrrr}
    \toprule
    Model & Arm & Audit reject (\%) & Coverage (\%) & Detect (\%) & Tests/task & Safe/call & Dup./call & Kills/1K tok. \\
    \midrule
    Qwen2.5 & Plain & 6.27 & 76.57 & 74.28 & 1.14 & 1.194 & 0.153 & 3.119 \\
    Qwen2.5 & Placebo & 6.36 & 58.33 & 56.04 & 0.84 & 0.686 & 0.135 & 1.102 \\
    Qwen2.5 & Real & 6.62 & 60.63 & 58.82 & 0.89 & 0.738 & 0.136 & 1.170 \\
    Qwen3.5 & Plain & 2.57 & 84.42 & 82.85 & 1.38 & 1.515 & 0.537 & 3.753 \\
    Qwen3.5 & Placebo & 3.92 & 90.34 & 88.53 & 1.51 & 1.857 & 0.534 & 1.871 \\
    Qwen3.5 & Real & 3.56 & 91.91 & 90.22 & 1.54 & 1.847 & 0.541 & 1.877 \\
    \bottomrule
  \end{tabular}}
\end{table}

The iterative arms have nearly identical token cost. Placebo and Real use
2.098 and 2.103 million attributed tokens for Qwen2.5, and 2.166 and 2.170
million for Qwen3.5. Per direction-round, Qwen2.5 Placebo and Real produce
0.827 and 0.874 reference-valid inputs, and 0.794 and 0.837 panel-confirmed
inputs. Qwen3.5 produces 1.917 and 1.906 reference-valid inputs, and 1.882 and
1.874 confirmed inputs. This motivates the matched replay below.

The frozen arms had equal nominal slots but not equal realized candidate yield.
Across task repeats and cross-fit directions, Qwen2.5 Real and Placebo produced
0.837 and 0.794 panel-confirmed reference-valid candidates per round; Qwen3.5
produced 1.874 and 1.882.  We replayed all 828 saved model--task--seed rows and
first required exact reproduction of both frozen oracle views.  We then retained
the earliest pairwise-minimum candidate count in each direction and round.

\begin{table}[h]
  \caption{E16-B post-hoc yield-matched Real--Placebo sensitivity (\pp, 95\% CI).}
  \label{tab:e16-yield-matched-appendix}
  \centering
  \small
  \begin{tabular}{lll}
    \toprule
    Matching policy & Qwen2.5 & Qwen3.5 \\
    \midrule
    Reference-valid & $-0.30$ [$-0.90$,$+0.30$] & $+0.78$ [$-0.62$,$+2.19$] \\
    Panel-safe/disconfirmed strata & $-0.35$ [$-0.91$,$+0.20$] & $+0.13$ [$-1.03$,$+1.23$] \\
    \bottomrule
  \end{tabular}
\end{table}

This replay makes no model calls or program executions and leaves the frozen
E16-B decision unchanged.  It is off-policy: later candidates were generated
from the full original histories, not from the truncated replay.  The result
therefore establishes sensitivity to observed candidate opportunity, not a
causal direct effect of semantic attribution.

\section{Human Semantic Audit}
\label{app:semantic-audit}

Two software engineering doctoral students independently reviewed the 300 cases
in separate randomized blind orders. Both received the same frozen annotation
schema and substantive instructions. For each case, a reviewer saw the problem
statement, public examples, and candidate input, but not the task identity,
generation method, archived output, program behavior, fault detections, panel
outcome, or the other reviewer's judgment. All 300 final rows from each reviewer
passed the same schema validator. Reviewer identities are withheld and replaced
with A and B in Table~\ref{tab:semantic-confusion} because direction is not
substantively important.

\begin{table}[h]
  \caption{Five-class judgment confusion matrix.  The two unused classes,
  ambiguous specification and insufficient information, have zero rows and
  columns and are omitted for space.}
  \label{tab:semantic-confusion}
  \centering
  \small
  \begin{tabular}{lrrr}
    \toprule
    & \multicolumn{3}{c}{Judge B} \\
    \cmidrule(lr){2-4}
    Judge A & Valid & Out of domain & Malformed \\
    \midrule
    Valid & 18 & 0 & 4 \\
    Out of domain & 1 & 195 & 0 \\
    Malformed & 2 & 18 & 62 \\
    \bottomrule
  \end{tabular}
\end{table}

The raw five-class exact agreement is $275/300=91.67\%$ and Cohen's
$\kappa=.823$.  After collapsing both invalid labels, agreement is
$293/300=97.67\%$ with $\kappa=.825$.  These values measure inter-rater
agreement between the two human reviewers; they do not establish that a shared
interpretation is semantically correct.

The specification-adjudication packet contains 67 unique cases. Its categories
are not mutually exclusive at selection time: 25 have inter-rater disagreement,
18 have reviewer consensus on validity and therefore need expected-output
review, 24 are a deterministic 10\% quality sample from otherwise unflagged
consensus cases, and four of the jointly valid cases are resource-limit-
sensitive. The union is 67 because those four overlap the valid-output category.
The packet is reserved for consensus or third-reviewer adjudication, which would
record a final validity label, output when applicable, notes, and whether the
evidence came from specification review, constraint checking, brute force, or
execution replay. Until then these cases are not force-resolved; the main
analysis uses the two frozen human ratings and preserves the audit-disconfirmed
terminology.

\section{Protocol Boundaries}
\label{app:boundaries}

The external protocol's three claim checks concern the existence of the oracle
problem, positive gain inflation, and non-superiority to matched resampling
within a $+2$\pp tolerance.  All passed for both model conditions.  The protocol
explicitly forbids using the external result to tune a generator or choose an
adapter.  The development fault-family analysis is exploratory, and its
confidence intervals are descriptive.  No cohort supports a causal model-scale
claim or a semantic frequency claim about source-level bug categories.  E14 is a prospectively locked extension on the
already exposed external cohort, not a new untouched replication.  E15 is a
post-hoc diagnostic.  Its real-minus-placebo contrast identifies no detectable
increment from fine-grained attribution under the constructed placebo, but it
does not establish equivalence or eliminate the possible value of aggregate
execution signals.
E16-A is the pristine qualification attempt and failed before model calls.
E16-B is outcome-blind with respect to model results but follows a disclosed
qualification amendment; its two inconclusive decisions neither confirm
equivalence nor establish practically important superiority.

\end{document}